\documentclass[%
aip,jcp,amsmath,amssymb, reprint, linenumbers
]{revtex4-2}
\usepackage{graphicx}
\usepackage{dcolumn}
\usepackage{bm}
\usepackage{amssymb}
\usepackage{amsmath}
\usepackage{supertabular}
\usepackage[table,x11names]{xcolor}
\usepackage{longtable}
\usepackage{colortbl}
\usepackage{multirow}\usepackage{braket}
\usepackage{esvect}
\usepackage{gensymb}
\usepackage{soul}
\let\oldAA\AA
\renewcommand{\AA}{\text{\normalfont\oldAA}}

\usepackage{qtree}
\usepackage{tipa}
\usepackage{appendix}
\usepackage{ulem}
\usepackage{xcolor}
\usepackage{tabularx,booktabs}
\usepackage{simplewick}
\usepackage{simpler-wick}
\usepackage{multirow, makecell}
\usepackage{scalerel}
\usepackage[framemethod=tikz]{mdframed}
\usepackage{mathrsfs}
\usepackage[colorinlistoftodos]{todonotes}
\presetkeys{todonotes}{inline,backgroundcolor=white}{}

\begin{document}
\nolinenumbers
\title{Efficient quantum state preparation through seniority driven operator selection}
\author{Dipanjali Halder$^{1}$, Dibyendu Mondal$^{1}$, Rahul Maitra$^{1,2,\dagger}$}
\affiliation{$^1$ Department of Chemistry, \\ Indian Institute of Technology Bombay, Powai, Mumbai 400076, India \\
$^2$ Centre of Excellence in Quantum Information, Computing, Science and Technology, \\Indian Institute of Technology Bombay, Powai, Mumbai 400076, India \\
$^\dagger$ rmaitra@chem.iitb.ac.in}


\begin {abstract}
Quantum algorithms require accurate representations of electronic states on a quantum device, yet the approximation of electronic wave functions for strongly correlated systems remains a profound theoretical challenge, with existing methods struggling to balance the competing demands of chemical accuracy and gate efficiency. Moreover, a critical limitation of the most of the state-of-the-art methods developed to date lies in their substantial reliance on extensive pre-circuit measurements, which introduce significant overheads and contribute to inefficiencies in practical implementation. To address these interconnected challenges and establish a harmonious synergy between them, we propose an algorithmic framework that focuses on efficiently capturing the molecular strong correlation through an ordered set of computationally less demanding rank-one and seniority-zero excitations, yielding a parameterized ansatz with shallow gate depth. 
Furthermore, to achieve minimal pre-circuit measurement overhead, we implement a selective pruning of excitations through a hybrid approach that combines intuition-based selection with shallow-depth, rank-one excitations driven uni-parameter circuit optimization strategy. 
With the incorporation of qubit-based excitations via particle-preserving exchange circuits, we demonstrate a further reduction in quantum complexities, enhancing the overall resource efficiency of the approach. With a range of challenging applications on strongly correlated systems, we demonstrate that our dynamic ansatz not only significantly enhances computational efficiency but also delivers exceptional accuracy, robustness, and resilience to the noisy environments inherent in near-term quantum hardware.
 
\end{abstract}

\maketitle
\section{Introduction}\label{introduction}
Solving the electronic Schrödinger equation lies at the heart of theoretical chemistry, offering critical insights into molecular structure, behavior, and interactions. Traditional classical methods, however, are severely limited by the curse of dimensionality, as the computational cost grows exponentially with the system size. Quantum computing offers a compelling alternative  to address these exponential bottlenecks~\cite{RevModPhys.92.015003, PhysRevLett.83.5162}, though its practical realization remains constrained by short coherence times, gate imperfections, and limited circuit depths.
To navigate these limitations, quantum-classical hybrid algorithms, notably the Variational Quantum Eigensolver (VQE)\cite{Peruzzo_2014} have emerged as a promising pathway toward achieving \textit{quantum advantage} in quantum chemistry. The feasibility of such hybrid algorithms on the near term quantum devices, however, relies on the ability of approximate wavefunction ansatzes to produce accurate energies and properties while minimizing the quantum complexities. 
Generally, the wavefunction ansatzes are typically constructed using either hardware-efficient or chemically motivated fermionic operators, each with distinct trade-offs\cite{Kandala2017, D1CS00932J, TILLY20221}. Hardware efficient ansatz, specifically designed to optimize the use of available quantum resources, offers a more pragmatic approach but fails to preserve physical symmetries and faces scalability limitations due to optimization challenges. Whereas the chemically motivated ansatzes, such as those based on traditional unitary coupled cluster (UCC) method or other problem-specific methods, inherently preserve physical symmetries, enable more straightforward optimization, and allow systematic improvement through higher-order excitations. However, they tend to be more resource-intensive, often demanding significantly deeper and complex circuits. Hence, the challenge lies in balancing the ansatz's capability and flexibility to capture complex many-electron correlations with the inherent limitations of near-term quantum hardware.\\ 
In addressing this challenge, significant progress has been made toward developing efficient and physically meaningful ansatzes. Among the earliest and most notable approaches is the unitary coupled cluster singles and doubles (UCCSD) method~\cite{Peruzzo_2014, Romero_2019}, which had set a foundational framework for further advancements in the field. However, due to the non-commuting nature of the operator components within UCCSD, the Trotterized implementation of this method is inherently non-unique, leading to substantial variations in accuracy across different operator orderings~\cite{Grimsley2020}. Moreover, as the complexity of molecular correlations escalates, Trotterized UCCSD may fail to achieve chemical accuracy, often incorporating numerous unnecessary excitations that do not contribute meaningfully to the correlation, yet significantly inflate the circuit depth, thereby exacerbating the challenges of efficient quantum simulation. Recent investigations have extended UCCSD by incorporating operators beyond conventional particle-hole excitations~\cite{doi:10.1021/acs.jctc.8b01004, doi:10.1063/5.0114688, 10.1063/5.0104815}. Notable among these advancements is k-UpCCGSD~\cite{doi:10.1021/acs.jctc.8b01004}, which achieves linear scaling by employing paired generalized rank-two excitations combined with generalized single excitations. As quantum systems grow in size and complexity, fixed ansatzes like k-UpCCGSD often fail to maintain the delicate balance between accuracy and resource efficiency. Consequently, it becomes imperative to develop dynamic approaches that can effectively tailor the ansatz as per system-specific criteria, ensuring both accuracy and computational feasibility. A prominent example of such a state-of-the-art approach is ADAPT-VQE~\cite{Grimsley_2019}, which efficiently extracts correlation energy using a dynamically identified, problem specific minimal subset of operators. However, its reliance on gradient-based operator selection introduces substantial measurement overhead, which scales unfavorably with the system size. Unlike the ADAPT-VQE approach, an alternative and notable approach in this context is the DISCretely Optimized Variational Quantum Eigensolver (DISCO VQE)~\cite{disco_vqe} which introduces a distinct strategy by incorporating products of spin symmetry-preserving fermionic operators, while simultaneously optimizing both the sequence of operators and their amplitudes. Addressing the high measurement overhead associated with adaptive algorithms, the present authors have introduced COMPASS-VQE~\cite{10.1063/5.0153182}, a protocol that identifies dominant operators using parallel one- and two-parameter circuits, thus reducing resource demands. The same set of authors have subsequently introduced two distinct ansatz compactification strategies: one grounded in first-principles-based many-body perturbation theory (COMPACT-VQE)~\cite{10.1063/5.0198277} and the other driven by unsupervised machine learning techniques (RBM-VQE)~\cite{D3SC05807G}. For further insights and related advancements in this area, the reader is referred to the literatures cited herein\cite{doi:10.1021/acs.jctc.0c00421, Ryabinkin2018,  Ryabinkin2020, fan2023circuitdepth, doi:10.1063/1.5141835, PhysRevResearch.2.033421, Fedorov2022unitaryselective, Mehendale2023, doi:10.1021/acs.jctc.9b00963, doi:10.1021/acs.jpca.3c01753, Haidar_2025, Feniou2023}.\\
Despite these advancements, the accurate representation of electronic wavefunctions at low computational cost remains a major challenge in the field, highlighting the importance of continued development of efficient and scalable methods. While highly accurate approaches often rely on the inclusion of intricate and computationally prohibitive higher-rank excitations, it is well established that the number of two-qubit CNOT gates required to implement an $n$-tuple fermionic or qubit excitation scales exponentially with $n$\cite{doi:10.1021/acs.jctc.2c01016}. Given that two-qubit gates, such as CNOTs, are significantly more error-prone (often by an order of magnitude) than the single-qubit gates, they inevitably constitute the primary source of gate errors in practical quantum hardware. Consequently, reducing the number of CNOT gates becomes imperative for realizing accurate and experimentally feasible quantum simulations. A conceivable route to address this challenge lies in the strategic parametrization of the electronic wavefunction, where the use of computationally efficient low-rank excitations may suffice to capture essential correlation effects, thereby circumventing the need for explicitly incorporating higher-rank excitations.  
In consonance with this idea, we propose an algorithm that relies predominantly on the computationally less demanding rank-one excitations, which serve as pivotal elements capable of spanning higher seniority sectors of the Hilbert space when supplemented by a sparse subset of rank-two, seniority-preserving paired excitations, thereby achieving an optimal balance between computational tractability and accurate description of strong electron correlations. In addition to  the use of structurally simple and computationally efficient operator pool, our method dynamically identifies the most significant and chemically relevant operators from the pool, simultaneously determining their relative ordering through a chemical intuition driven  shallow-depth, uni-parameter circuit optimization strategy. We term this approach as Seniority-informed Unitary Ranking and Guided Evolution based Variational Quantum Eigensolver (SURGE-VQE).\\
In the following section, we outline the key theoretical concepts underlying our method, SURGE-VQE. A detailed, step-by-step description of the ansatz construction workflow is presented in Sec. \ref{pipeline}. Sec. \ref{complexity} provides a detailed analysis of the associated quantum complexities. In Sec. \ref{results}, we summarize the results from our study of ground state energetics, followed by Sec. \ref{conclusion}, which concludes with a discussion of our findings and outlines possible future research directions.



\section{Theory}\label{theory}
\subsection{VQE-UCC Framework}\label{VQE-UCC}
The time-independent Schrödinger equation for an $n-$electron system, under the Born–Oppenheimer approximation, is mathematically expressed via the eigenvalue equation, as follows:
\begin{eqnarray}
    H\ket{\Psi}=E\ket{\Psi}
\end{eqnarray}
where $E$ and $\ket{\Psi}$ are the electronic ground state energy and wavefunction respectively, and $H$ refers to the electronic Hamiltonian, which under the second quantized formulation leads to 
\begin{eqnarray}
H=\sum\limits_{pq}h_{pq}a_{p}^{\dagger}a_{q}+\frac{1}{4}\sum\limits_{pqrs}h_{pqrs}a_{p}^{\dagger}a_{q}^{\dagger}a_{s}a_{r}.
\end{eqnarray}
where $h_{pq}$ and $h_{pqrs}$ correspond to one- and two-electron integrals respectively. The indices $p,q,r$ and $s$ refer to the molecular spinorbitals. $a_{p}^\dagger$ and $a_{s}$ corresponds to the creation and annihilation operators that add and remove electrons from spinorbitals p and s respectively.\\
In the Variational Quantum Eigensolver (VQE) algorithm, a central component is the parametrized unitary operator, which is employed to prepare the wavefunction ansatz, $\ket{\Psi(\vec{\theta})}$.
\begin{eqnarray}
    \ket{\Psi(\vec{\theta})}={U(\vec{\theta})}\ket{\Psi_{0}}.
\end{eqnarray}
A common and usual choice for the reference $\ket{\Psi_{0}}$ in the quantum chemistry community is the Hartree-Fock reference, denoted by $\ket{\Phi_{HF}}$. The parametrized wavefunction is then variationally optimized following the Rayleigh-Ritz variational principle.
\begin{eqnarray}
    E_{0}=\underset{\vec{\theta}}{min}\bra{\Psi_{0}}U^{\dagger}(\vec{\theta})HU(\vec{\theta})\ket{\Psi_{0}} 
\label{cost_function}
\end{eqnarray}
A widely adopted choice for the unitary operator $U$ in the quantum chemistry community is the physically motivated unitary coupled cluster (UCC). The UCC wavefunction is defined as: 
\begin{eqnarray}
    \ket{\Psi_{UCC}}=e^{{T}-{T}^\dagger}\ket{\Phi_{HF}} = e^{{\tau}} \ket{\Phi_{HF}}
\end{eqnarray}
where $T$ in the exponent is the cluster operator that encapsulates the electron excitations (singles, doubles, triples etc.).
\begin{eqnarray}
    T &=& T_{1} + T_{2} + \hdots \\
    &=& \sum_{i;a}\theta_{ia}a_{a}^\dagger a_{i} + \sum_{i,j;a,b}\theta_{ijab}a_{a}^\dagger a_{b}^\dagger a_{i} a_{j} + \hdots
\end{eqnarray}
The subscripts $i,j$ refer to occupied spinorbitals and the indices $a, b$ refer to unoccupied spin orbitals. The coefficients $\theta_{ia}$ and $\theta_{ijab}$ denote the cluster amplitudes.\\
Since accounting for all possible excitations is computationally infeasible due to the exponential scaling with system size, the UCC expansion is usually truncated. The most common truncation occurs at the doubles level, leading to the Unitary Coupled Cluster Singles and Doubles (UCCSD). This approximation significantly reduces the complexity, requiring approximately $\mathcal{O}(N^{4})$ parameters, where $N$ represents the number of orbitals in the system.\\
Often, due to the fixed structure of the UCC ansatz, it fails to account for the specific chemical characteristics of the molecule under consideration. As a result, it inherently includes numerous excitations that may not be chemically relevant. These superfluous excitations can increase the computational cost (and hence the associated circuit complexities) without significantly improving the accuracy of the wavefunction, thus necessitating strategies for more efficient pruning or judicious selection of chemically relevant excitations. Notably, a significant subset of these relevant excitations stems from higher-order excitation operators, which, while essential for improving accuracy, are accompanied by a dramatic escalation in the quantum gate complexity, leading to circuits of prohibitively large depth. As a result, there is a pressing need for the development of methodologies that enable efficient quantum state preparation  while maintaining a low-depth quantum circuit. 

\subsection{A Road-map toward a Shallow Depth Ansatz}\label{choice_ops}
As a computationally efficient and sufficiently precise framework for characterization of static correlation effects, seniority-preserving many-body methodologies have consistently retained a prominent and instrumental role in the theoretical arsenal.  A particularly noteworthy advancement in this regard has been the conceptualization of pair coupled cluster doubles (pCCD)\cite{Tecmer2014, PhysRevB.89.201106, 10.1063/1.4880819}, which introduces a significant simplification to the conventional coupled cluster wavefunction by constraining the allowed excitations to paired ones, where an electron pair is excited from a single spatial orbital to a distinct spatial orbital, thereby ensuring that the resulting wavefunction consists exclusively of seniority-zero (or seniority-preserving) determinants. The pCCD wavefunction can be written as follows:
\begin{eqnarray}
    \ket{\Psi_{pCCD}}=e^{{\hat{T}}}\ket{\Phi_{HF}}
\end{eqnarray}
where $\ket{\Phi_{HF}}$ refers to the closed shell reference determinant i.e., the Hartree-Fock reference and $\hat{T}$ is restricted to the paired excitation operators,
\begin{eqnarray}
    \hat{T}=\sum_{i,a}{\theta_{ia}}a_{a}^{\dagger}a_{\bar{a}}^{\dagger}a_{\bar{i}}a_{i}
\end{eqnarray}
where i and a refers to the indices for occupied and virtual orbitals and the coefficients $\theta_{ia}$ refer to the cluster amplitudes.\\
The term seniority is defined in relation to the number of broken electron pairs and is quantified by the number of unpaired electrons or singly occupied spatial orbitals within a Slater determinant. Seniority number serves as a measure of the degree to which electron pairing has been disrupted, with a higher seniority indicating a greater number of unpaired electrons in the Slater determinant\cite{10.1063/1.4880819, henderson}. The representation of the wavefunction through low-seniority sectors has proven highly efficient in capturing the correlation energy, enabling a more compact and computationally tractable description of the electronic structure. Motivated by the seniority-driven partitioning of the Hilbert space, we adopt a similar strategy, wherein we employ sparse pair-coupled cluster double excitation operators to span the lowest seniority sectors of the Hilbert space. However, the restriction to seniority zero sectors can, in certain cases, be overly stringent; particularly those exhibiting strong correlation effects, inherently require the inclusion of higher seniority sectors to accurately capture the full complexity of their electronic structure. Consequently, in addition to the seniority-preserving rank-two operators, we also incorporate generalized rank-one operators, thereby broadening the formalism to encompass higher seniority sectors and ensuring a more complete and accurate description of the system’s electronic correlations. It is imperative to note, at this juncture, that the adoption of this specific pair of operators for describing the many-electron ground state wavefunction is by no means a novel or unprecedented approach, as the same set of operators has previously been employed in a variety of methodologies, as exemplified by works such as kUpCCGSD\cite{doi:10.1021/acs.jctc.8b01004}, DISCO-VQE\cite{disco_vqe} and tUPS~\cite{tups}, which have been driven by this very operator selection.  While the profound kinship between these methodologies and our formalism is evident—particularly with respect to the shared pool of operators—it is equally critical to underscore that our approach distinguishes itself through the introduction of strategically designed, intuition-driven measures, as well as dynamic reference-based operators sequencing strategies, which enable a more effective pruning of operators and a more precise ordering, thereby setting our method apart and ensuring its enhanced efficacy in capturing the system’s electronic structure.\\
An alternative rationale for the selection of operators within the context of the wavefunction description through our formalism can be construed as a direct corollary arising from two widely recognized and well-established observations: 
\begin{enumerate}
    \item{As articulated by Evangelista and co-workers, an arbitrary $n-$electron operator can be decomposed into nested commutators of rank-one and rank-two excitation operators~\cite{10.1063/1.5133059}. For example, a rank-three operator, denoted by $\hat{\tau}_{ijk}^{abc}$ can be written as follows:
    \begin{equation}
           \hat{\tau}_{ijk}^{abc}\rightarrow [\hat{\tau}_{ij}^{ae},[\hat{\tau}_{m}^{e}, \hat{\tau}_{mk}^{bc}]]
    \end{equation}}
    \item{As demonstrated by Burton and co-workers, any arbitrary two-electron unpaired excitation can be decomposed into nested commutators encompassing seniority preserving paired excitations and generalized one-electron operators~\cite{disco_vqe}. For example an unpaired two-electron excitation, $\hat{\tau}_{i\bar{j}}^{a\bar{b}}$ can be written as follows:  
    \begin{equation}
           \hat{\tau}_{i\bar{j}}^{a\bar{b}}\rightarrow [[\hat{\tau}_{i\bar{i}}^{a\bar{a}},\hat{\tau}_{\bar{j}}^{\bar{i}}], \hat{\tau}_{\bar{a}}^{\bar{b}}]
    \end{equation}}
\end{enumerate}

Having established the foundational basis for operator selection, the next step lies in identifying the chemically \textit{``relevant"} subset among them and determining their relative ordering - an algorithmic approach for which is outlined in the following subsection.

\subsection{Ansatz construction workflow}\label{pipeline}
The step-by-step framework of SURGE-VQE is outlined below, organized into four core components, each addressing a distinct stage of the ansatz construction protocol.
\begin{enumerate}
    \item {\textbf{Intuition-driven operator pooling:} The operator pool, $\mathscr{O}$ is characterized by paired two-electron excitations ($T_{2,p}$) and generalized one-electron excitations ($T_{1,g}$).
    \begin{equation}
        \mathscr{O} = \{T_{2,p}, T_{1,g}\}
    \end{equation}}
    While both excitation types are crucial for accurately describing the electronic structure, the operator pool is further streamlined to enhance efficiency and ensure physical relevance. Guided by orbital symmetry constraints, this refinement retains only those generalized one-electron excitations that occur between orbitals within the same irreducible representation of the molecular point group.
    \item {\textbf{Dynamic seniority preserving reference tailoring:} } With Hartree-Fock as the conventional reference state, each paired two-electron excitation operator from the pool $\mathscr{O}$ is employed to tailor an ensemble of seniority preserving reference states, which are optimized through a one-parameter VQE approach incorporating the usual UCC structure. Notably, the number of seniority-preserving reference states generated aligns exactly with the cardinality ($n_o n_{v}$) of the paired excitation operators constituting the pool. Although the Hartree-Fock determinant itself is a seniority-preserving (or seniority-zero) determinant, in this manuscript, we explicitly do not refer to it as such unless specifically mentioned. Instead, we use the term ``seniority-preserving" or ``seniority-zero" solely for determinants constructed via the action of paired rank-two excitation operators upon the Hartree-Fock reference.\\ 
    The state,  $\ket{\Psi^{\alpha}}$ generated via $\alpha$-th paired excitation can be mathematically written as,
    \begin{equation}
        \ket{\Psi^{\alpha}} =e^{\theta^{\alpha}_{*}\tau_{2,p}^{\alpha}}|\phi_{HF}\rangle, \alpha \epsilon [1,n_{o}n_{v}]
    \end{equation}
    where $\tau^{\alpha}_{2,p}$ refers to the anti-hermitian sum of $\alpha$-th paired excitation and de-excitation and the parameter $\theta^{\alpha}_{*}$ corresponds to the optimal parameter obtained from uni-parameter optimization of the following energy functional, 
    \begin{equation}
        E^{\alpha} = \min_{\mathbf{\theta^{\alpha}}}\langle \Phi_{HF}|e^{-\theta^{\alpha}\tau_{2,p}^{\alpha}}\hat{H}e^{\theta^{\alpha}\tau_{2,p}^{\alpha}}|\Phi_{HF}\rangle, 
    \end{equation}
    Once the optimal energies corresponding to these seniority-preserving reference states are evaluated, we proceed to compute the energy deviation relative to the Hartree-Fock reference energy, denoted as $\Delta E^{\alpha}$
    \begin{equation}
        \Delta E^{\alpha} = E_{HF} - E^{\alpha} 
    \end{equation}
    The resulting energy deviations subsequently govern the precise ordering of the operators, a critical aspect that profoundly influences the accuracy and reliability of the ansatz as a whole.\\
    Please note that the deviation $\Delta E^\alpha$ and the optimal state $\ket{\Psi^\alpha}$ can be  efficiently determined with massive parallelization if multiple quantum devices are available. However, the reliance on quantum hardware for this step is not imperative, as the same objective can be attained through classical approximations, such as many-body perturbative methods like MP2, which offer a viable alternative rooted in well-established theoretical frameworks.

    \item{\textbf{Operator blocking and inter-block reshuffling:} } The seniority-preserving paired excitations are regarded as the initiating elements in the ordering of operators. The energy deviations, calculated for each paired excitation in the preceding step serve as the criterion for their arrangement within distinct operator blocks. The ensemble of $n_{o}n_{v}$ paired operators corresponding to energy deviations ordered as $\Delta E^{\beta} > \Delta E^{\alpha} > \hdots > \Delta E^{\gamma} $ is allocated to $n_{o}n_{v}$ distinct operator blocks. These blocks are assigned in such a manner that the excitation with the highest energy deviation, $\tau_{2,p}^{\beta}$, is allocated to the first block—namely, the one that acts directly upon the Hartree-Fock state while the subsequent excitations are allocated to subsequent blocks in descending order of their respective energy deviations, with $\tau_{2,p}^{\gamma}$ occupying the final block. 
    \begin{eqnarray}
    \ket{\Psi}&=&\big[e^{\tau_{2,p}^{\gamma}}\big]_{n_{o}n_{v}}\hdots\big[e^{\tau_{2,p}^{\alpha}}\big]_{2}\big[e^{\tau_{2,p}^{\beta}}\big]_{1}\ket{\Phi_{HF}}
    \end{eqnarray}
    The terms in parentheses denote the block operators, and the subscripts $1$, $2$, $\hdots$ , $n_{o}n_{v}$ denote the block numbers.\\
    Following the initial formation and energy deviation-based reshuffling of these operator blocks, we proceed to expand each block by appending additional sub-blocks, thereby accommodating the singles in their respective sub-blocks.
    \item{\textbf{Sub-block prescreening, block augmenting and intra-block reshuffling:} } In order to effectively incorporate singles into our ansatz, we undertake a selective screening of the most dominant and significant single excitations, reapplying the uni-parameter Variational Quantum Eigensolver (VQE) approach.
    However, rather than depending on the conventional Hartree-Fock reference state, we adopt the seniority-zero reference states to effectively capture the important single excitations and impose an ordering within each block. With $\ket{\Psi^{\alpha}}$ as the reference,  we systematically construct uni-parameter quantum circuits corresponding to each of the singles (denoted by I) selected from the pruned operator pool. These circuits are subsequently subjected to variational optimization, enabling the assessment of the relative significance and hierarchical arrangement of the singles within the prescribed operator blocks. Denoting the optimal energy as $E^{\alpha I}$ it can be expanded mathematically as shown below:
    \begin{equation}
    E^{\alpha I} = \min_{\mathbf{\theta^{I}}}\langle \Psi^{\alpha}|e^{-\theta^{I}\tau_{1,g}^{I}}\hat{H}e^{\theta^{I}\tau_{1,g}^{I}}|\Psi^{\alpha}\rangle,
    \end{equation}
    where $I \epsilon [1,N^{2})$, $N$ refers to the total number of spin-orbitals.\\
    The energy deviation value $\Delta E^{\alpha I} = E^{\alpha} -  E^{\alpha I}$ then serves as the additional prescreening criterion for effective singles (or sub-block) within a block, allowing selection only if the following condition is met:
    \begin{equation}
        \Delta E^{\alpha I} > \epsilon
    \end{equation}
    where $\epsilon$ is an user defined threshold.\\
    Based on the values of $\Delta E^{\alpha I}$, the pruned singles (or sub-blocks) are reshuffled to determine a near-optimal ordering within that block, such that the singles with the least energy deviation are placed at the beginning of the block. The final ansatz is then denoted by Eq. \ref{final_ansatz}
    \begin{eqnarray}
        \ket{\Psi} &=& \big[ \hdots e^{\tau_{1,g}^{I}} e^{\tau_{1,g}^{J}} e^{\tau_{2,p}^{\gamma}}\big]_{n_{o}n_{v}}  
        \hdots \big[e^{\tau_{2,p}^{\alpha}}\big]_{2}\nonumber \\ 
        && \big[\hdots e^{\tau_{1,g}^{J}} e^{\tau_{1,g}^{K}} e^{\tau_{1,g}^{I}} e^{\tau_{2,p}^{\beta}}\big]_{1} \ket{\Phi_{HF}}\nonumber\\
        &=& \prod_{\alpha=1}^{n_{o}n_{v}}\bigg[\bigg[\prod_{I=1}^{M_{\alpha}} e^{\tau_{1,g}^{I}}\bigg]e^{\tau_{2,p}^{\alpha}}\bigg]\ket{\Phi_{HF}}, M_{\alpha} \leq N^{2} 
    \label{final_ansatz}
\end{eqnarray}
where $M_{\alpha}$ refers to the total number of sub-blocks or singles corresponding to block $\alpha$. And the indices $I$, $J$ and $K$ refer to the singles. Their ordering within each block  is guided by the energy deviation values. For example, if we consider block 1, the singles or (sub-blocks) follows the following (energy-deviation) order:
\begin{eqnarray}
\hdots < \Delta E^{\beta J} < \Delta E^{\beta K} < \Delta E^{\beta I}.    
\end{eqnarray}
To reduce the measurement and optimization complexity further for sub-block prescreening and ordering, we apply uni-parameter optimization based pruning only to a single spin sector, as spin-complementary singles typically yield the same result. While singles from the other spin sector are included in the final ansatz, their selection and ordering are dictated by their spin-complementary counterparts. This step effectively reduces the sub-block search space to half of the total number of singles.
\end{enumerate}

\subsection{Circuit Complexity and Measurement Bounds:}\label{complexity}
As evident from Eq.\ref{final_ansatz}, the final ansatz generated by SURGE-VQE comprises only seniority preserving paired rank-two excitations, which exhibit a known quadratic scaling of $\mathcal{O}(n_o n_v)$, along with dynamically selected rank-one generalized operators. Notably, while a given generalized rank-one operator cannot appear more than once within a block, it may reappear in different blocks, allowing for selective repetition of single excitation operators. The inclusion of rank-one generalized operators introduces an upper-bounded scaling of $\mathcal{O}(N^{2})$, where $N=n_{o}+n_{v}$. Consequently, the overall parameter scaling is governed by the leading contribution from generalized single excitations, ensuring that the scaling remains upper-bounded at $\mathcal{O}(N^{2})$. Incorporating the Jordan-Wigner overhead of $\mathcal{O}(N)$ further leads to an approximate total gate count of $\mathcal{O}(N^{3})$. It is worth emphasizing at this point that although SURGE-VQE comes from a different conceptual standpoint, it shares significant structural resemblance with the k-UpCCGSD ansatz. This is primarily because the operator pool employed in our formulation constitutes a subset of the k-UpCCGSD operator pool. Hence, leveraging the analysis provided by Lee and co-workers\cite{doi:10.1021/acs.jctc.8b01004}, one may infer that the corresponding quantum circuit for our ansatz can be expressed at worst by a linear gate depth, $\mathcal{O}(N)$. \\
An inherent advantage of our algorithm lies in its prescreening and operator reshuffling methodology, which is meticulously orchestrated through a shallow-depth, single-parameter circuit-based optimization. This strategic approach ensures an exceptionally low measurement complexity per prescreening cycle, scaling merely as $\mathcal{O}(N^{4})$ (due to the quartic number of Hamiltonian terms in second quantization). In contrast, conventional adaptive schemes exhibit a progressive escalation in measurement overhead, where the initial complexity of $\mathcal{O}(N^4)$ in the first iteration undergoes an iterative amplification, ultimately reaching $\mathcal{O}(N^4 N_{p})$ in the final iteration, with $N_{p}$ denoting the total number of parameters incorporated into the ansatz. Additionally, adaptive approaches require gradient evaluations at each iteration, necessitating the measurement of $\mathcal{O}(N^4 N_o)$ observables per iteration where $\mathcal{O}(N_o)$ denotes the size of the operator pool.\\

\begin{figure*}
    \centering
    \includegraphics[width=1\textwidth]{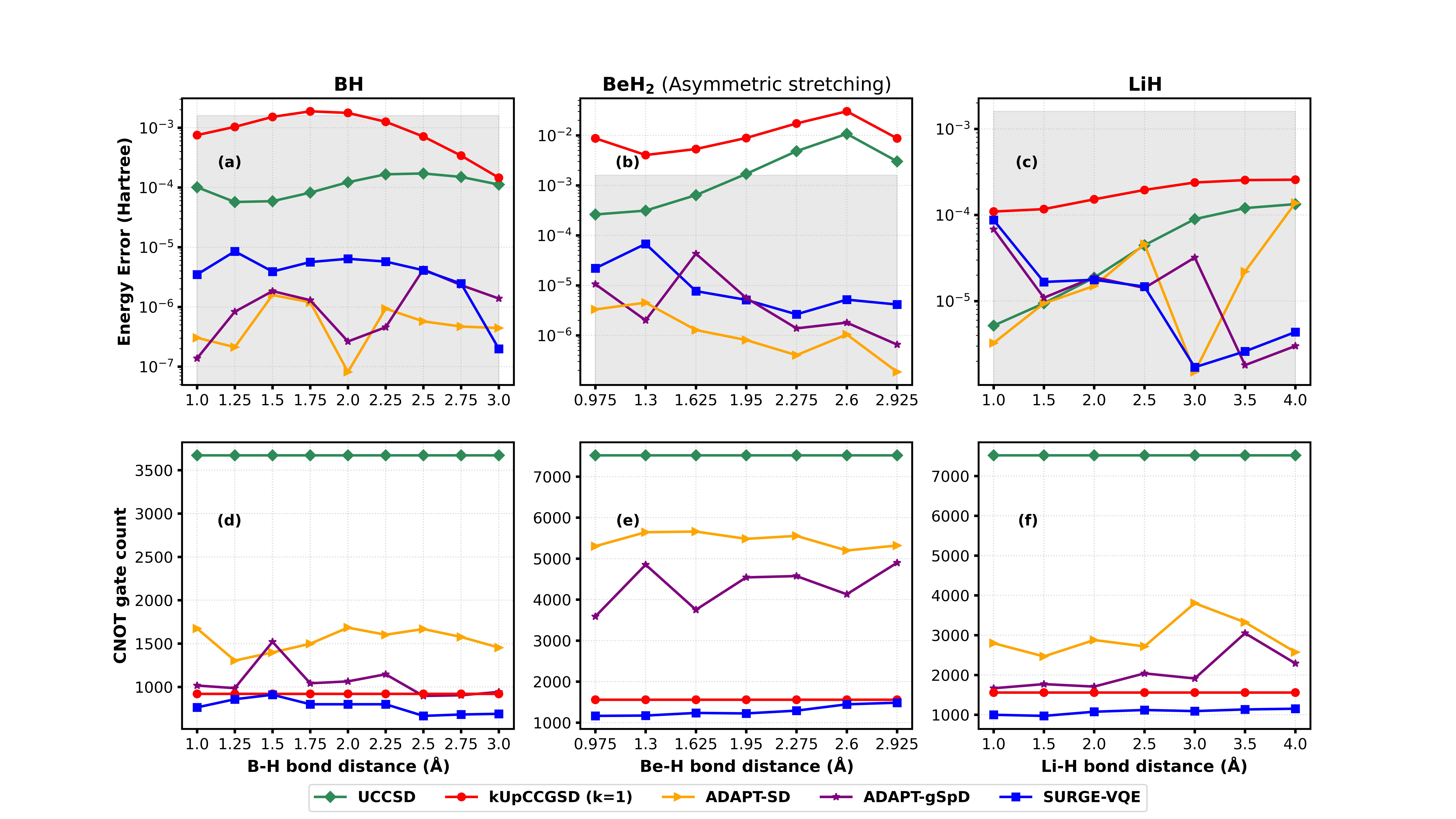}
    \caption{\textbf{Energy error relative to FCI (first row) and CNOT gate count per function evaluation (second row) across the potential energy profiles of $BH$ (first column), $BeH_{2}$ (second column), and $LiH$ (third column).}}
    \label{fig:statevector_pes}
\end{figure*}

\begin{figure}
\centering
\begin{minipage}{0.9\columnwidth}
    \includegraphics[width=\linewidth]{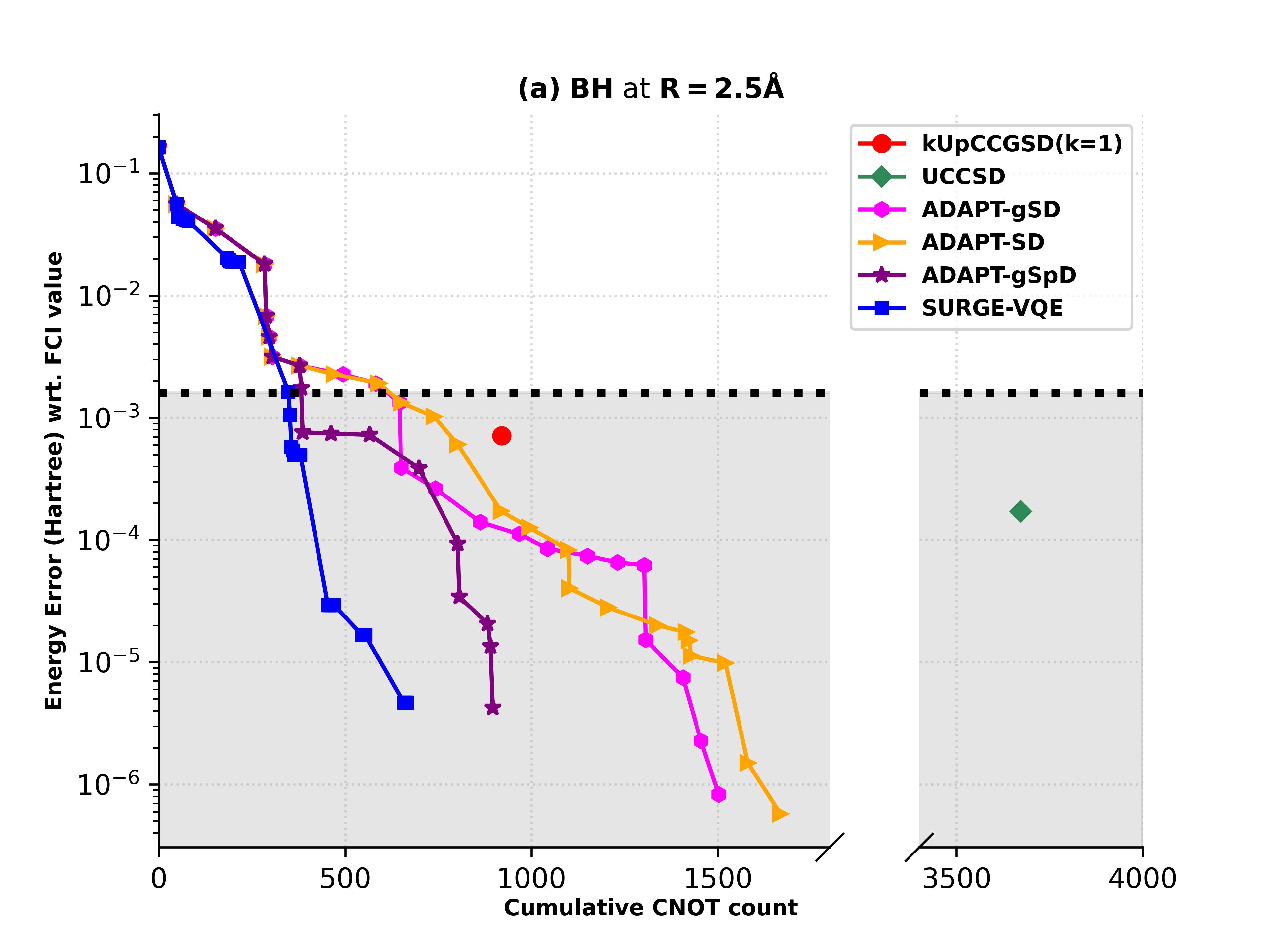}
    \label{fig:cnot_bh}
    \vspace{0.3cm} 
    \includegraphics[width=\linewidth]{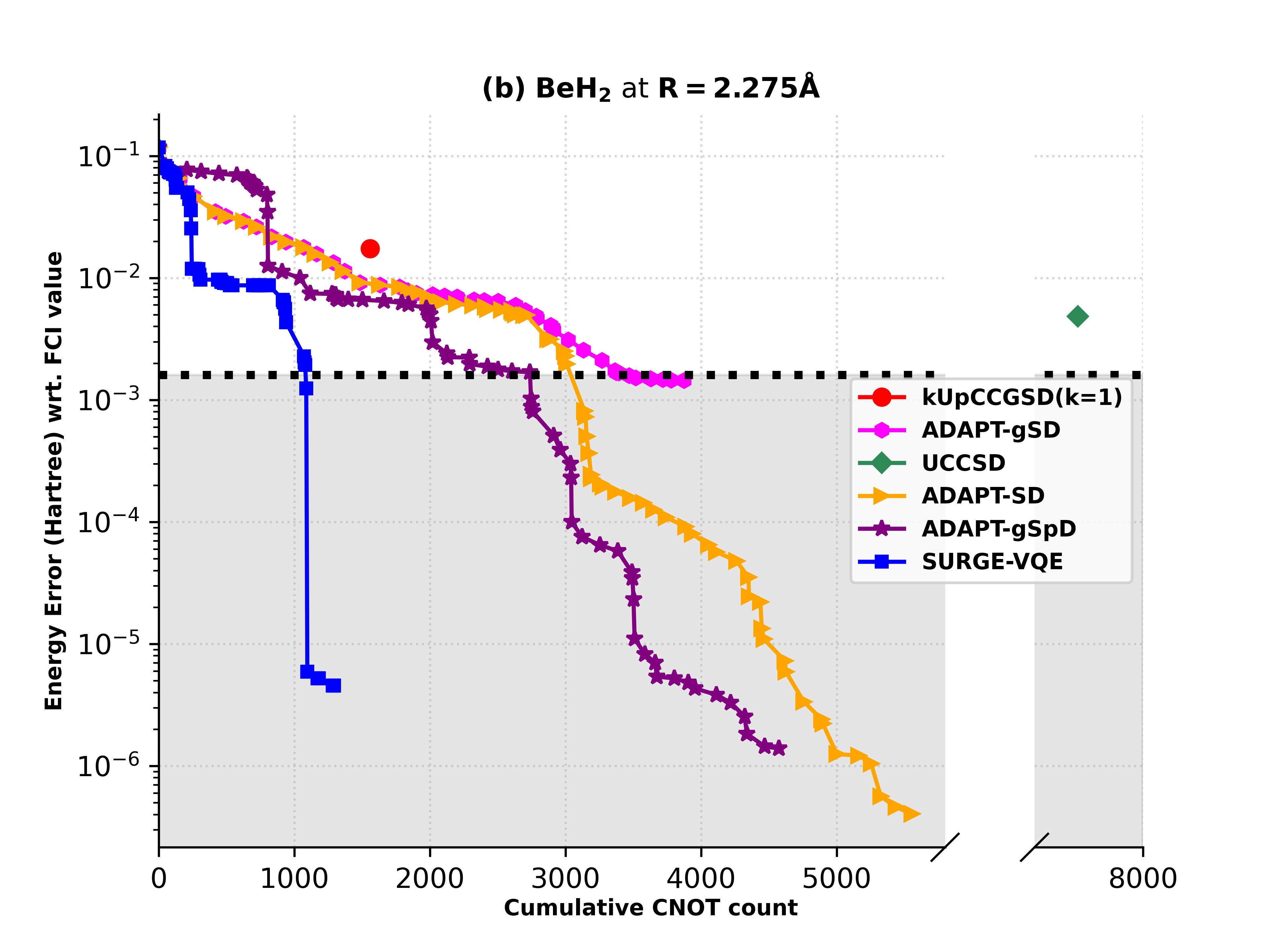}
    \label{fig:cnot_beh2}
    \vspace{0.3cm}
    \includegraphics[width=\linewidth]{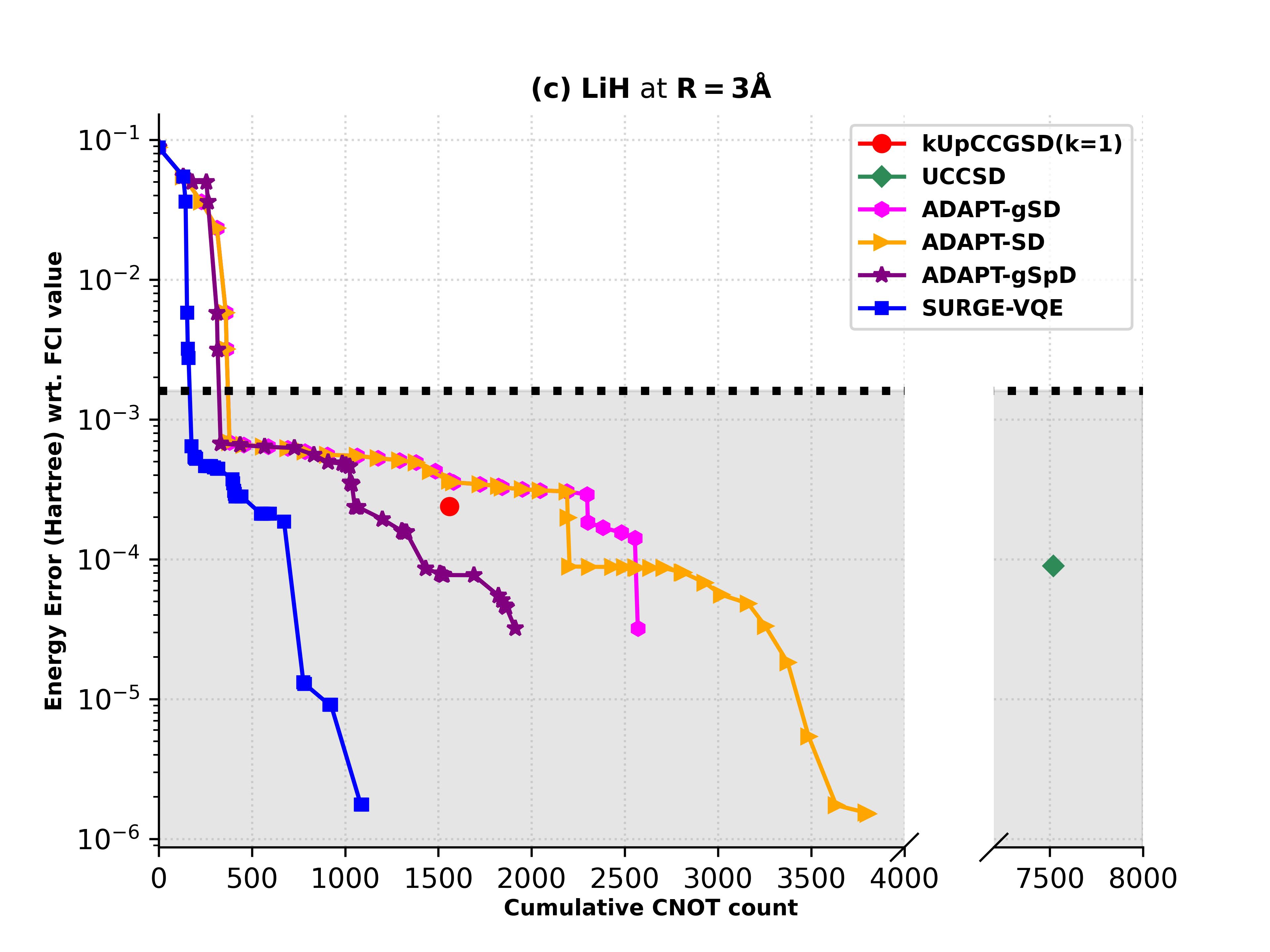}
    \caption{\textbf{Sequential depiction of energy error relative to FCI as a function of the cumulative two-qubit CNOT gate count for (a) $BH$ at bond distance, $R=2.5\AA$, (b)$BeH_{2}$ at $Be-H$ bond distances, $R=2.275\AA$ and $0.9*R=2.0475\AA$, (c) $LiH$ at bond distance, $R=3\AA$. Each point on the graph corresponds to the incremental addition of a single operator to the ansatz, indicating how the gradual growth of the ansatz (one parameter at a time) affects both energy accuracy and overall circuit complexity.}}
    \label{fig:cnot_lih}
\end{minipage}
\end{figure}

\section{Results and Discussions}\label{results}
\subsection{Computational Methodology}
We have demonstrated the performance of SURGE-VQE by numerically simulating the ground state potential energy curves for three representative molecular systems: $BH$, $LiH$ and linear $BeH_{2}$ (with asymmetric stretching) in the minimal STO-3G\cite{doi:10.1063/1.1672392} basis set. All the numerical calculations have been carried out using an internally developed Python code, leveraging the PySCF\cite{https://doi.org/10.1002/wcms.1340} and Qiskit Nature\cite{the_qiskit_nature_developers_and_contrib_2023_7828768} packages. We have adopted the Jordan Wigner transformation scheme\cite{doi:10.1063/1.4768229} for fermion to qubit mapping and the gradient based L-BFGS-B\cite{doi:10.1137/0916069} optimization algorithm for VQE simulations on the statevector simulator. The variational parameters have been set to zero at initialization. And the sub-block prescreening threshold, $\epsilon$ for our algorithm have been set to $10^{-6}$ for all the illustrated molecular test cases. All the ADAPT-VQE calculations have been carried out using the code provided by Qiskit. 
\subsection{Potential energy surfaces and two-qubit CNOT gate count as functions of nuclear co-ordinates}
In this section, we comprehensively elucidate the robustness of our approach by specifically emphasizing its capability to yield highly accurate results, while concurrently demonstrating its efficacy in terms of computational efficiency. To this end, we have employed three challenging molecular systems - bond stretching in $BH$ and $LiH$, and asymmetric stretching in $BeH_{2}$ - to rigorously assess the performance of SURGE-VQE. These systems exhibit pronounced electronic correlation effects, which are known to pose significant difficulties for traditional computational methods, thereby offering a rigorous benchmark for assessing the robustness of our approach. In order to 
further highlight the strengths and limitations of SURGE-VQE in comparison to established methods, we have performed the same set of calculations using a range of conventional and state-of-the-art techniques. These include the traditional UCCSD, kUpCCGSD, ADAPT-VQE with the singles and doubles operator pool (denoted as ADAPT-SD), and ADAPT-VQE with an operator pool identical to ours, namely generalized singles and paired doubles (denoted as ADAPT-gSpD). For the kUpCCGSD, we have employed the simplest version with minimal gate complexity (i.e, k=1), and for the ADAPT-SD and ADAPT-gSpD calculations, we have set both the eigenvalue threshold and the gradient threshold to $10^{-8}$. Fig.\ref{fig:statevector_pes} showcases the relative performance of our ansatz across two different metrics: accuracy or energy error relative to the classically exact FCI energy and two-qubit CNOT gate count per function evaluation, offering a detailed comparison of its effectiveness in the context of these key computational factors.\\
We begin our analysis by investigating the dissociation energy profile of the $BH$ molecule. For this analysis, the simulations have been conducted using the STO-3G basis set, keeping the core orbitals frozen throughout the computational procedure. With the core orbitals frozen, the molecule effectively reduces to a 10 qubit (spinorbital), 4 electron system. For bond lengths ranging from $1 \AA$ to  $3 \AA$, we have performed variational optimization employing our proposed ansatz, alongside other established ansätze, to rigorously compare their performance in capturing the dissociation behavior of the molecule. As evident from Fig.\ref{fig:statevector_pes}(a), SURGE-VQE consistently achieves chemical accuracy (defined as an error within 1.0 kcal/mol, or approximately 1.6 $mE_{h}$ relative to the FCI energy) across the entire potential energy curve. The energy deviation from the classically exact FCI ranges between 8.5 $\mu E_{h}$ at $R=1.25\AA$ and 0.199 $\mu E_{h}$ at $R=3\AA$. As expected, the accuracy of our approach is orders of magnitude better than the UCCSD and kUpCCGSD $(k=1)$ ansätze. However, it is noteworthy that the accuracy of the ADAPT variants (especially ADAPT-SD) surpasses that of our approach by a considerable margin. One of the primary factors contributing to the superior performance of the ADAPT variants is the meticulous choice of gradient and eigenvalue thresholds, which have been set at relatively low and generous values. But it is important to note that such high accuracy is not without its trade-offs. Specifically, this enhanced performance comes at the expense of significantly higher two-qubit CNOT gate requirements, as clearly demonstrated in Fig.\ref{fig:statevector_pes}(d). It is worth emphasizing that the CNOT gate requirements for ADAPT-SD are an order of magnitude greater than those of SURGE-VQE, thus highlighting the considerable gate depth incurred in the pursuit of higher accuracy within the ADAPT framework.\\
Next, we examine a more challenging test case for our analysis: the asymmetric stretching of the linear $BeH_{2}$ molecule. Taking $R$ as the variable, we set one Be–H bond distance to $R$, while the other is fixed at 0.9$\times$$R$ thus leading to an asymmetric scenario. With the core orbitals frozen and within the confines of the STO-3G basis set, $BeH_{2}$ can be identified as a 12 qubit (spinorbital) and 4 electron system. As indicated by Fig.\ref{fig:statevector_pes}(b), SURGE-VQE yields consistent accuracy across the entire potential energy surface, outperforming both UCCSD and kUpCCGSD. While kUpCCGSD fails to achieve chemical accuracy throughout the surface, UCCSD also falls short of chemical accuracy at the stretched geometries. In contrast, SURGE-VQE not only achieves chemical accuracy but also maintains precision within a few microhartree, even in the challenging regime of stretched geometries. Expectedly, ADAPT-SD again demonstrates superior performance, exhibiting an accuracy that is an order of magnitude better, though this comes at the expense of a substantial escalation in quantum complexity as demonstrated in Fig.\ref{fig:statevector_pes}(e).  \\
The last system considered in this study is $LiH$, which is frequently used as a critical benchmark for evaluating the performance of newly developed electronic structure methods. In 
STO-3G basis set, $LiH$ represents a 12 qubit (spinorbital), 4 electron system. A comprehensive comparative study, detailing the energy error relative to the FCI and the associated quantum complexities as a function of the $Li–H$ bond length, is presented in Fig.\ref{fig:statevector_pes}(c) and \ref{fig:statevector_pes}(f) respectively. From the analysis presented in Fig.\ref{fig:statevector_pes}(c), it can be concluded that SURGE-VQE initially achieves an energy error on the order of tens of microhartree for the starting geometries or bond lengths. However, as the bond lengths are stretched, the energy deviations improve markedly, falling within the microhartree range. Notably, this enhancement is realized  with a striking reduction in quantum complexities which is evident from Fig.\ref{fig:statevector_pes}(f), emphasizing the efficiency of our method across varying bond lengths. As anticipated, SURGE-VQE surpasses both kUpCCGSD and UCCSD in terms of accuracy (except in the case of initial geometries for UCCSD) while simultaneously maintaining an extraordinarily minimal quantum complexity. Furthermore, it is of particular note that SURGE-VQE follows a trajectory that closely aligns with the ADAPT-gSpD curve, while simultaneously distinguishing itself through an extraordinarily low requirement for two-qubit CNOT gates during execution. For instance, at $R=2.5\AA$, both ansätze achieve nearly identical accuracies. However, the gate count for our ansatz is 1120, while that for ADAPT-gSpD is approximately double, reaching 2024. In stark contrast, the ADAPT-SD variant fails to demonstrate consistent improvements in accuracy across the entire potential energy surface. While it exhibits favorable performance near equilibrium, its accuracy begins to degrade significantly as the system moves away from this region. Notably, in the domains corresponding to strong molecular correlations, the CNOT gate count for ADAPT-SD escalates to levels that are nearly three times, or even more, than that required by SURGE-VQE, further highlighting the need for more gate-efficient alternatives to ADAPT-SD in such regions.\\
It is important to note that our proposed ansatz may not reach the same level of variational expressiveness as ADAPT family of methods, which are known for their accuracy and flexibility. Rather, our goal is to propose a resource-conscious alternative that maintains reasonable accuracy while significantly reducing quantum gate overhead. In this regard, our ansatz strikes a practical balance - delivering chemically relevant accuracy at a low and manageable gate cost, making it a highly gate-efficient option for near-term quantum applications.

\subsection{Energy Landscape study}
In this section, we examine the energy convergence trajectory as error relative to the full configuration interaction (FCI) as a function of the cumulative two-qubit CNOT gate count. After prescreening the operators and establishing their ordering according to the proposed strategy, we now sequentially expand the ansatz, starting from the Hartree-Fock reference state, incorporating one operator at a time to explore the energy landscape. The wavefunction at $k$-th step of the expansion can be expressed as in Eq. \ref{growth_eqn}. 
\begin{eqnarray}
    \ket{\Psi^{(k)}(\vec{\theta}^{(k)})} = e^{{\theta^{(k)}\tau^{(k)}}}\ket{\Psi^{(k-1)}(\vec{\theta}^{(k-1)})}
    \label{growth_eqn}
\end{eqnarray}
This way the ansatz is constructed as a growth ansatz where at the $k$-th iterative step, the operator at the $k$-th position of the original ordered operator pool is added and by construction, unlike ADAPT-VQE, this approach bypasses any requirement of gradient measurement to choose the operator from the pool. 
At each step of this ansatz expansion, a variational optimization is performed, with the initial parameter ($\theta^{(k)}$) value for the newly added operator ($\tau^{(k)}$) set to zero, while the previously added parameters ($\vec{\theta}^{(k-1)}$) are initialized to their optimized values obtained at the previous step. The energy at $k-$th expansion step is thus given by Eq. \ref{growth_energy}
\begin{eqnarray}
    E^{(k)}=\min_{\vec{\theta}^{(k)}}\langle \Psi^{(k)}{(\vec{\theta}^{(k)})}|\hat{H}| \Psi^{(k)}{(\vec{\theta}^{(k)})}\rangle
    \label{growth_energy}
\end{eqnarray}
As mentioned previously, the sequential wavefunction expansion and operator addition adhere to the ordering defined in Eq. \ref{final_ansatz}. Hence, based on the pre-defined ordering, the form of $\tau^{(k)}$ is as follows:  
\begin{center}
\[
\tau^{(k)} = \begin{cases}
\tau^{\alpha}_{2,p}, & \text{if k-th operator is a paired double} \\
\tau^{I}_{1,g}, & \text{if k-th operator is a generalized single.} 
\end{cases}
\]
\end{center}
Fig.\ref{fig:cnot_lih} illustrates how the energy evolves as the prescreened operators are sequentially incorporated into the ansatz across all the three test cases: $BH$, $BeH_{2}$ (under asymmetric stretching) and $LiH$- evaluated at different bond lengths as indicated in the respective plots. Rather than presenting operator counts, we focus on the cumulative CNOT gate count along the x-axis, offering a hardware-relevant metric that directly reflects the quantum execution cost, thus providing a more practical perspective than parameter-based comparisons.\\
We have also included results for various ADAPT-VQE variants alongside our proposed method for comparison. We have plotted the energy deviations relative to FCI as a function of the cumulative CNOT gate count, with each point corresponding to the addition of an operator in successive ADAPT iterations for ADAPT-SD and ADAPT-gSpD (both evaluated using eigenvalue and gradient thresholds set to $10^{-8}$). Additionally, we have also presented results for the ADAPT-VQE employing the generalized singles and doubles (gSD) operator pool, hereafter referred to as ADAPT-gSD. Given the substantially larger size of the gSD pool relative to the gSpD and SD pools, we have adopted a more relaxed eigenvalue threshold of $10^{-6}$, while maintaining the gradient threshold at $10^{-8}$, in order to reduce computational resource demands that would otherwise escalate with a stricter eigenvalue threshold.\\
The blue trajectory corresponding to our method reveals a characteristic \textit{`burrowing'} behavior in the energy-CNOT landscape - an early and steep descent in energy error followed by a rapid approach toward  improved accuracy with minimal incremental cost in CNOT gates. Unlike other methods that tend to plateau or descend more gradually, our ansatz effectively \textit{`burrows'} through the energy landscape before other methods begin to significantly lower their errors.\\
Notably, the burrowing behavior also suggests that our method shows heuristic evidence of avoiding redundant or energetically inefficient directions in Hilbert space, instead steering toward optimal directions that refines the wavefunction with higher precision. It reflects a kind of quantum resource frugality - squeezing out more accuracy per CNOT, which is vital for practical implementation on NISQ-era devices where noise and decoherence severely limit circuit depth.\\
While it is important to acknowledge that ADAPT-based ansätze achieve exceptional energy accuracy, our objective is fundamentally different. Rather than optimizing purely for accuracy, our focus lies in striking an effective balance between energy precision and circuit resource demands. Consequently, we do not aim for the absolute lowest energy error, but instead prioritize achieving reasonably accurate results with significantly reduced gate depth - an essential consideration for practical deployment on NISQ-era quantum hardware.

\begin{figure*}
    \centering
    \includegraphics[width=1\textwidth]{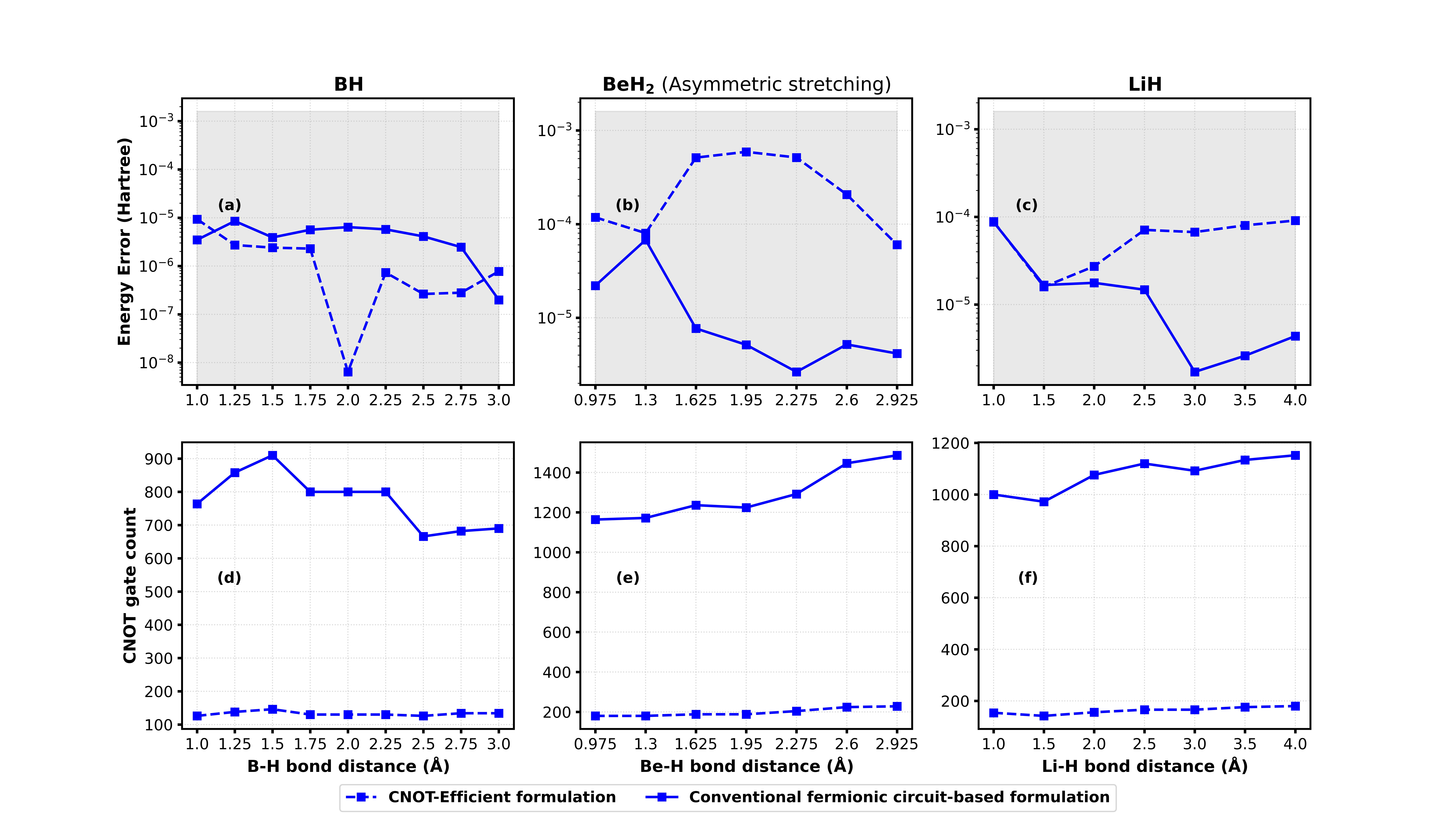}
    \caption{\textbf{Comparative study of energy errors relative to FCI (first row) and CNOT gate count per function evaluation (second row) across the potential energy profiles of $BH$ (first column), $BeH_{2}$ (second column), and $LiH$ (third column), utilizing both conventional fermionic excitation-based circuit and the CNOT-efficient formulation of our proposed ansatz.}}
    \label{fig:pes_cnot}
\end{figure*}

\begin{figure*}
    \centering
    \includegraphics[width=16cm, height=9cm]{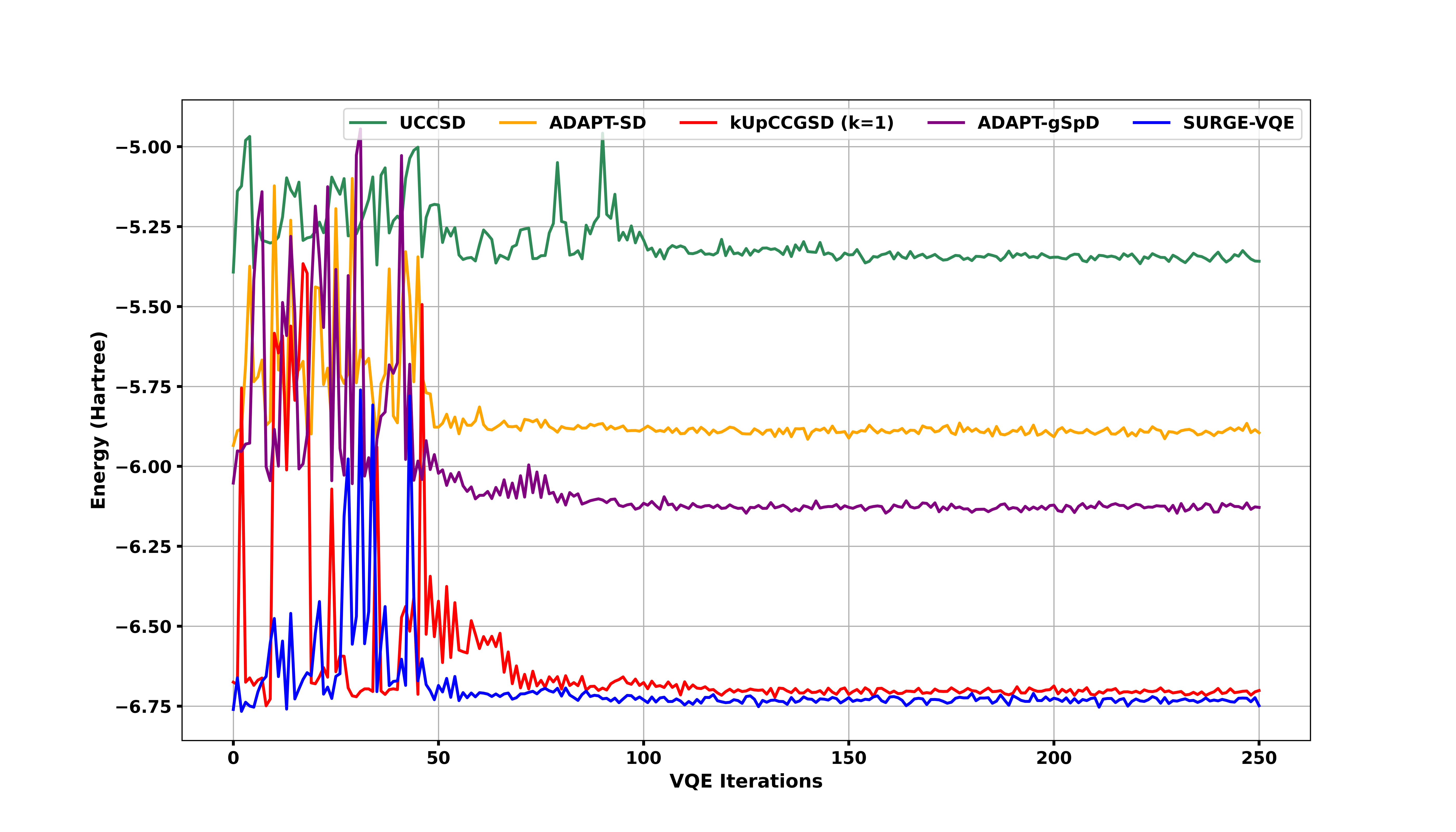}
    \caption{\textbf{Comparison of energy vs. VQE iterations under noise profile extracted from IBM's \textit{FakeMelbourne} for $LiH$ at $R=3\AA$}}
    \label{fig:noise_lih}
\end{figure*}
\subsection{CNOT Efficient implementation}
A fundamental challenge in NISQ computing lies in the ability to implement a sufficient number of entangling gates with high fidelity, as imperfections in gate operations significantly impact the accuracy and feasibility of quantum algorithms. Consequently, minimizing the use of entangling gates is crucial for the practical realization of variational quantum algorithms on near-term quantum hardware. Traditional circuit constructions for fermionic operators rely on CNOT staircases, an approach that, while theoretically sound, proves highly inefficient in the context of current hardware constraints. While the method hitherto presented reduces the CNOT gate count considerably while maintaining high accuracy, we further optimize the CNOT gate usage, in this subsection by replacing the fermionic operators with simpler qubit operators $Q_i ^\dagger = {(X_i - iY_i)}/{2}$ and  $Q_i = {(X_i + iY_i)}/{2}$. These qubit excitations  are implemented via particle-preserving exchange circuits, as introduced by Yordanov et al\cite{PhysRevA.102.062612, Yordanov2021}. The gate complexity corresponding to such a circuit implementation of a specific qubit operator is known to be $\mathcal{O}(1)$.\\
In this section, we evaluate the performance of the final ansatz generated via SURGE-VQE with such a CNOT-efficient formulation\cite{doi:10.1021/acs.jctc.2c01016, Xia_2021, https://doi.org/10.1002/qua.27001} and compare it against the standard fermionic-excitation based circuit implementation. The comparison is conducted in terms of both the energy error relative to FCI and the two-qubit CNOT gate count across the molecular potential energy surfaces of all three test cases.

For all three molecular systems discussed in this work, the CNOT-efficient variant consistently demonstrates a notable reduction in entangling gate counts as evident from Fig.\ref{fig:pes_cnot}. In particular, for $LiH$ and $BeH_2$, the reduction is nearly an order of magnitude compared to the conventional fermionic circuit-based formulation. Interestingly, in the case of $BH$, the CNOT-efficient approach not only achieves a lower gate count but also yields slightly better accuracy than its fermionic counterpart. However, this trend does not hold for $BeH_{2}$ and $LiH$, where a slight decline in accuracy is observed. This outcome is not at all counter-intuitive, as the removal of parity terms is naturally expected to influence the overall accuracy. While this trade-off introduces a mild compromise in energy precision, the resulting errors remain  within chemically meaningful limits.

\subsection{Performance under noisy circumstances}
In this section, we delve into the performance of our ansatz when executed on a noisy backend, analyzing its robustness and behavior in the presence of quantum noise.  To simulate realistic noise effects, we have incorporated the noise model from IBM's FakeMelbourne backend to study the impact on ground-state energy calculations using our method.

Figure \ref{fig:noise_lih} presents the ground state energy vs. VQE iterations plot for LiH at $3\AA$ bond distance, comparing our approach against conventional methods such as UCCSD, kUpCCGSD (k=1), ADAPT-SD, and ADAPT-gSpD under backend-induced noise (without any error mitigation). For the variational optimization, we employed the COBYLA optimizer with a maximum iteration limit set to 250 in all cases, initializing all parameters at zero. Each curve represents an average data obtained from 40 independent runs. Also we performed noisy simulations using 10,000 shots to ensure statistical reliability.\\  
SURGE-VQE demonstrates better inherent accuracy relative to other approaches, primarily due to its significantly lower gate count. It is important to note that this study emphasizes the inherent robustness of different ansatze under hardware noise, rather than their absolute accuracy. To achieve chemical accuracy, one must employ appropriate error mitigation protocols. However, given the inherently low depth of our ansatz, the additional cost associated with such mitigation protocols is generally expected to be lower in comparison to other methods, and hence SURGE-VQE may be considered as a potential candidate to achieve chemical accuracy in near-term quantum hardware when implemented in conjunction with suitable error mitigation strategies. More importantly, the superiority in robustness and accuracy achieved by this theory is maintained when the simulations are performed in an ideal noiseless scenario as discussed previously, and thus the method is expected to be equally beneficial to the quantum computing community when the hardware quality improves in the coming years.

\section{Conclusion and future directions}\label{conclusion}
In this study, we have presented an algorithmic framework for efficient quantum state preparation, leveraging a structured hierarchy of computationally less demanding rank-one and seniority-zero excitations. With our algorithm, we have focused on reducing the quantum complexities (or execution gate depth to be specific) via three key aspects: first by selecting an operator pool composed of gate efficient operators; second, by intuitively excluding the irrelevant operators from the pool; and third, by dynamic reference-based uni-parameter optimization driven operator selection. This approach yields a parameterized ansatz with inherently shallow (linear or at worst, sub-quadratic) gate depth. Moreover, the incorporation of qubit excitations through particle-preserving exchange circuits enables an additional reduction in gate count, bringing it down to under 180 CNOTs even for challenging cases such as $LiH$ at highly stretched bond lengths, while maintaining energy precision within tens of microhartree.\\

To further minimize pre-circuit measurement overhead, we have employed a selective pruning strategy that blends intuitive orbital symmetry driven selection with uni-parameter circuit optimization approach - aimed not only at identifying the chemically relevant or effective operators but also at determining their relative ordering to improve accuracy. Please note that the uni-parameter optimization strategies come with both advantages and limitations. One significant advantage is the substantial reduction in measurement overhead. Since only a single parameter needs to be optimized at a time, the optimization process typically requires very few iterations (significantly fewer than multi-parameter optimization). Additionally, each individual parameter optimization can be run independently in parallel quantum architectures. However, this approach may not be sufficient for molecules with strong multireference signature. Hence, we do not claim that the relative ordering of operators derived from our method is necessarily the best or most optimal. Rather, we emphasize that the algorithm we developed are heuristic in nature, and their effectiveness depends on the specific context of the system being studied.\\
Looking forward, we are interested in integrating amplitude reordering strategies\cite{Lan2022} to achieve better operator ordering and enhanced accuracy. Another promising direction involves reducing the pre-circuit measurement overhead in the ansatz construction pipeline by employing more chemically intuitive approaches, which would help mitigate the imperfections introduced by the noisy quantum devices. In addition, we plan to integrate the adiabatic decoupling framework, augmented with non-iterative auxiliary subspace corrections\cite{10.1063/5.0210854, 10.1063/5.0229137} into our dynamic SURGE-VQE algorithm to enable further resource optimization. 
Last but not the least, it will be worth studying the performance of SURGE-VQE (with the CNOT-efficient formulation) on real quantum hardware, along with suitable mitigation routines to better understand its practical applicability and limitations.

\section*{Acknowledgements}
D.H. thanks the Industrial Research and Consultancy Center (IRCC), IIT Bombay and D.M. acknowledges the Prime Minister’s Research Fellowship (PMRF), Government of India for their research fellowships. R.M. thanks Anusandhan National Research Foundation (ANRF) (erstwhile SERB), Government of India (Grant Number: MTR/2023/001306).

\section*{Conflict of Interests}
The authors have no conflict of interests to disclose.
\section*{Data Availability}
The data that support the findings of this study are available from the corresponding author upon reasonable request.

\bibliography{literature}
\end{document}